\documentclass[aps,prb,twocolumn,showpacs,superscriptaddress]{revtex4-1}

\usepackage{graphicx}
\usepackage{amsmath}
\usepackage{multirow}
\usepackage{tabularx}
\usepackage{color}
\usepackage[colorlinks,linkcolor=blue,anchorcolor=blue,citecolor=blue]{hyperref}

\begin{document}

\title{Effect of iron vacancies on the magnetic order and spin dynamics of the spin ladder BaFe$_{2-\delta}$S$_{1.5}$Se$_{1.5}$}	
\author{Zengjia Liu*}
\author{Xiao-Sheng Ni}
\email{The authors contributed equally to this work.}
\author{Lisi Li}
\author{Hualei Sun}
\author{Feixiang Liang}
\affiliation{Center for Neutron Science and Technology, School of Physics, Sun Yat-Sen University, Guangzhou, 510275, China}
\author{Benjamin A. Frandsen}
\affiliation{Department of Physics and Astronomy, Brigham Young University, Provo, Utah 84602, USA}
\author{Andrew D. Christianson}
\author{Clarina dela Cruz}
\affiliation{Materials Science and Technology Division, Oak Ridge National Laboratory, Oak Ridge, Tennessee 37831, USA}
\author{Zhijun Xu}
\affiliation{NIST Center for Neutron Research, National Institute of Standards and Technology, Gaithersburg, Maryland 20899, USA}
\affiliation{Department of Materials Science and Engineering, University of Maryland, College Park, Maryland 20742, USA }
\author{Dao-Xin Yao}
\affiliation{Center for Neutron Science and Technology, School of Physics, Sun Yat-Sen University, Guangzhou, 510275, China}
\author{Jeffrey W. Lynn}
\affiliation{NIST Center for Neutron Research, National Institute of Standards and Technology, Gaithersburg, Maryland 20899, USA}
\author{Robert J. Birgeneau}
\affiliation{Department of Physics, University of California, Berkeley, California 94720, USA }
\affiliation{Materials Science Division, Lawrence Berkeley National Laboratory, Berkeley, California 94720, USA}
\author{Kun Cao}
\email{caok7@mail.sysu.edu.cn}
\author{Meng Wang}
\email{wangmeng5@mail.sysu.edu.cn}
\affiliation{Center for Neutron Science and Technology, School of Physics, Sun Yat-Sen University, Guangzhou, 510275, China}

\begin{abstract}
Quasi-one-dimensional iron chalcogenides possess various magnetic states depending on the lattice distortion, electronic correlations, and presence of defects. We present neutron diffraction and inelastic neutron scattering experiments on the spin ladder compound BaFe$_{2-\delta}$S$_{1.5}$Se$_{1.5}$ with $\sim$6\% iron vacancies. The data reveal that long-range magnetic order is absent, while the characteristic magnetic excitations that correspond to both the stripe- and block-type antiferromagnetic correlations are observed. First-principles calculations support the existence of both stripe and block-type antiferromagnetic short-range order in the experimental sample. The disappearance of long-range magnetic order may be due to the competition between these two magnetic orders, which is greatly enhanced for a certain concentration of iron vacancies, which we calculate to be about 6\%, consistent with the measured iron vacancy concentration. Our results highlight how iron vacancies in the iron-based spin ladder system strongly influence the magnetic ground state.
\end{abstract}

\maketitle

\section{Introduction}

The mechanism of high-temperature superconductivity (SC) has remained a puzzle since the initial discovery of copper-based high-temperature superconductors~\cite{bednorz1986possible}. SC emerges for both the copper- and iron-based  families of superconductors upon the suppression of antiferromagnetic (AF) order through charge doping or pressure~\cite{kamihara2008iron,hsu2008superconductivity,lumsden2010magnetism,vignolle2007two,paglione2010high,johnston2010puzzle}. AF spin fluctuations have been put forth as an essential ingredient for high-temperature superconductivity~\cite{scalapino2012common}. Recently, pressure-induced superconductivity was found in the quasi-one-dimensional (1D) iron-based spin ladder compounds BaFe$_2X_3$ ( $X=$ S, Se), placing the 1D Fe spin ladder systems into the iron-based SC family~\cite{takahashi2015pressure,yamauchi2015pressure,ying2017interplay}. In contrast to the metallic parent compounds of the layered iron-based superconductors, both BaFe$_2$S$_3$ and BaFe$_2$Se$_3$ are insulating. It is speculated that the origin of superconductivity may be tied to a bandwidth-controlled Mott transition resulting from the increased transfer of Fe $3d$ electrons across the inter-ladder bonds. These findings have prompted extensive experimental and theoretical research to determine the role of magnetism in the superconducting spin ladder compounds and their connection to the layered cuprates and iron chalcogenide superconductors~\cite{stewart2011superconductivity,caron2012orbital,krzton2011synthesis,arita2015ab,dong2014bafe,mourigal2015block,hirata2015effects,ootsuki2015coexistence,suzuki2015first,Chi2016,patel2017pairing,zhang2018sequential,li2018phase}.

The two end members of the BaFe$_2$S$_{3-x}$Se$_x$ phase diagram possess distinct crystal and magnetic structures. BaFe$_2$S$_3$ crystallizes into an orthorhombic structure with space group $Cmcm$, as shown in Fig.~\ref{fig1}. The Fe atoms exhibit a stripe AF ground state~\cite{klepp1996mixed,wang2017strong,yu2020structural,mourigal2015block}. In contrast, BaFe$_2$Se$_3$ crystallizes a lower symmetry orthorhombic space group $Pnma$ in which the Fe ladders are slightly tilted~\cite{yu2020structural,mourigal2015block,wang2017strong,zheng2020room}. The Fe magnetic moments order into a block AF ground state, as shown in Fig.~\ref{fig1} (c). To understand the evolution of the magnetic ground state in BaFe$_2$S$_{3-x}$Se$_x$ as functions of chemical substitution and pressure close to superconductivity, systematic neutron diffraction investigations have been conducted under ambient~\cite{Wu2019,Wu2020,yu2020structural} and applied pressure\cite{Chi2016,Zheng2018}. The block AF order is robust up to 6.8 GPa at 120 K in BaFe$_2$Se$_3$~\cite{Wu2019}. With S substitution on Se sites, the block AF order survives until approximately a 1:2 ratio of Se to S, at which point stripe AF order appears~\cite{Du2019,yu2020structural}. Interestingly, short-range correlations of both the block and stripe type have been observed in the same sample of BaFe$_2$S$_2$Se, suggestive of a nontrivial intermediate state between the long-range-ordered versions of the two magnetic ground states~\cite{yu2020structural}.

Anderson localization induced by the S and Se substitution is suggested to be crucial for the appearance of superconductivity in the 1D ladder system~\cite{Sun2020}. Disorder also influences the magnetic properties. In particular, the magnetic order is highly sensitive to iron vacancies, as demonstrated by the spin-glass-like behavior observed in BaFe$_{1.8}$Se$_3$~\cite{Saparov2011} and the suppression of block-type long-range magnetic order by 4$\%$ iron vacancies in BaFe$_{2-\delta}$S$_{1.5}$Se$_{1.5}$ and 2$\%$ iron vacancies in BaFe$_{2-\delta}$S$_{0.67}$Se$_{2.33}$~\cite{yu2020structural}. The destruction of long-range magnetic order with such low concentrations of magnetic vacancies underscores the sensitive interplay between the magnetic interactions and detailed atomic structure in the Fe ladder system. To our knowledge, no deep understanding of the effect of iron vacancies on the magnetic ground state has been achieved, yet this may be related to the superconducting mechanism under high pressure. Hence, it is necessary to elucidate the role of iron vacancies in determining the magnetic ground state.

In this paper, we investigate the magnetic order and spin dynamics of BaFe$_{2-\delta}$S$_{1.5}$Se$_{1.5}$ with 6\% Fe vacancies. Neutron powder diffraction (NPD) measurements demonstrate that no long-range magnetic order develops, although short-range block AF correlations are observed at low temperature. Inelastic neutron scattering (INS) reveals the existence of both block- and stripe-type magnetic excitations, indicative of competition between these two types of magnetic orders. Using density functional theory (DFT), we calculated the magnetic ground state of BaFe$_{2-\delta}$S$_{1.5}$Se$_{1.5}$ for a range of iron vacancy concentrations. We find that a paramagnetic ground state can be ruled out for iron vacancy concentrations below 10\%. The calculated ground state for the vacancy-free compound is the block AF state, in agreement with experiments\cite{Du2019,yu2020structural}. However, the relative energy of the stripe AF state decreases with increasing iron vacancies, eventually becoming more energetically favorable than the block state for vacancy concentrations above $\sim$6\%. Combined with our experimental observations, we thus conclude that the magnetic ground state of our sample consists of both the stripe and block AF short-range orders. The disappearance of long-range magnetic order at such low concentration of iron vacancies appears to be caused by the competition between the two types of magnetic order, which is greatly enhanced with the introduction of iron vacancies.

\section{Experimental and theoretical methods}

Single crystals with a nominal composition of BaFe$_2$S$_{1.5}$Se$_{1.5}$ were grown by the Bridgman method\cite{Sun2020,yu2020structural}. A sample of single crystals weighing 2.55~g total were ground into powder for neutron scattering experiments. The NPD experiment was conducted on the HB-2A instrument at the High Flux Isotope Reactor, Oak Ridge National laboratory (HFIR, ORNL), using a monochromatic beam with a wavelength of $\lambda$ = 2.4105~\AA. The powder diffraction patterns were refined by the Rietveld method using the FullProf Suite\cite{rodriguez1993recent}.
Neutron diffraction measurements on single crystals were carried out on the BT7 thermal triple axis spectrometer at the NCNR\cite{Lynn2012}. A closed cycle refrigerator was used to control the sample temperature.
The INS experiment was carried out on the ARCS time-of-flight chopper spectrometer at the Spallation Neutron Source (SNS), ORNL. The measurements were conducted at 5~K with incident beam energies of E$_i$ = 20 and 150 meV. The corresponding energy resolutions were $\Delta E=1.0$ and 7.0 meV, respectively, as determined by the full width at half maximum of the energy cuts at $\Delta E=0$ meV. The $SpinW$ program was utilized to simulate INS spectra and compare with the experimental results\cite{Toth2015}. Magnetic susceptibility and resistivity measurements were performed using a physical property measurement system (PPMS, Quantum Design). Energy-dispersive X-ray spectroscopy (EDS) (EVO, Zeiss) was employed to determine the composition of the crystals.

\begin{figure}[t]
	\includegraphics[scale=0.33]{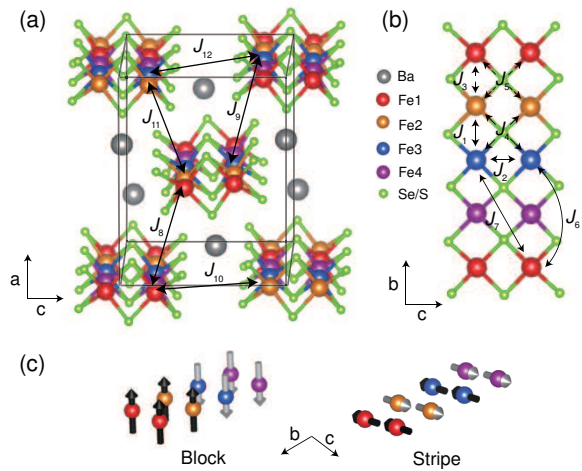}
	\caption{(a) Visualization of the crystal structure of BaFe$_2$S$_{1.5}$Se$_{1.5}$. (b) Structure of an individual Fe$X$ ($X=$ S/Se) ladder. $J_1\cdots J_{12}$ represent the magnetic exchange interactions between the marked magnetic ions. (c) Diagram of block and stripe antiferromagnetic orders. The directions of the magnetic moments are represented by the black and gray arrows.}
	\label{fig1}
\end{figure}

First-principles calculations were carried out with the Vienna Ab initio Simulation Package (VASP), based on density functional theory\cite{kresse1993ab,kresse1996efficient}. We used the Perdew-Burke-Ernzerhof functional with a spin-polarized generalized gradient approximation (GGA). The projector augmented-wave (PAW) \cite{blochl1994projector} method with a 500~eV plane-wave cutoff was used, and a $6\times4\times4$ Monkhorst-Pack $k$-point mesh allows the calculation to converge well. The spin-polarized GGA we used was combined with onsite Coulomb interactions, $U$, included for Fe 3$d$ orbitals (GGA $+ U$)\cite{liechtenstein1995density}. We employed $U = 1$ eV and $J = 0.4$ eV, which achieved values of the magnetic moments and band gap that are consistent with experiments. We started with the experimental atomic structure and then relaxed the crystal structure until the forces acting on each atom were less than 1 meV/\AA. In order to simulate the S substitution of Se, we employed the virtual crystal approximation (VCA) \cite{bellaiche2000virtual} for the same atomic coordinates of Se and S atoms.  The supercell we used is a monoclinic crystal cell with 32 Fe atoms, which can accommodate both the block and stripe AF orders. For the block AF order, the Fe1 and Fe2 spins share the same orientation, while the Fe3 and Fe4 spins are antiparallel to Fe1 and Fe2. For the stripe AF order, Fe1 and Fe4 spins share the same orientation, opposite to the Fe2 and Fe3 spins [see Fig. \ref{fig1} (c)]. The effect of iron vacancies was modeled by removing a given number of randomly selected Fe atoms from the supercell, calculating the results, and then averaging the results with those from equivalent calculations performed with the same number of iron atoms removed from other randomly selected positions.

\section{Results and discussions}

\begin{table}[b]
	\caption{\label{tab:table1} The refined structural parameters for the sample of nominal composition BaFe$_2$S$_{1.5}$Se$_{1.5}$ determined from NPD data at 4~K (goodness of fit $\chi^2=$ 19.5, $R_p=13.9$\%, $R_{wp}=$ 16.3\%, and  $R_{exp}=$3.7\%).}
	\begin{ruledtabular}
		\begin{tabular}{ccc}
			Crystal system&Orthorhombic\\
			Space group&$Pnma$\\
			Unit cell parameters&$a=11.5480(4)$\AA\\
			&$b=5.3208(1)$\AA\\
			&$c=8.9910(3)$\AA\\
			&$\alpha=\beta=\gamma=90^{\circ}$\\
			Atomic parameters&$(x,y,z)$&Occupancy\\
			Ba&$(0.1919,0.25,0.524)$&0.50\\
			Fe&$(0.4936,0.0017,0.3506)$&0.945(9)\\
			Se1&$(0.3530,0.25,0.233)$&0.22(2)\\
			Se2&$(0.623,0.25, 0.491)$&0.14(1)\\
			Se3&$(0.3999,0.25,0.8193)$&0.40(2)\\
			S1&$(0.3530,0.25,0.233)$&0.28(2)\\
			S2&$(0.623,0.25,0.491)$&0.36(1)\\
			S3&$(0.3999,0.25,0.8193)$&0.10(2)\\
		\end{tabular}
	\end{ruledtabular}
\end{table}

\begin{figure}[t]
	\centering
	\includegraphics[scale=0.5]{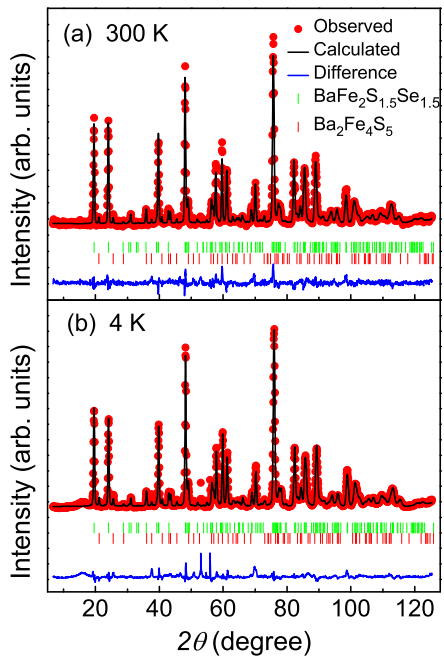}
	\caption{Observed and calculated neutron powder diffraction patterns for BaFe$_2$S$_{1.5}$Se$_{1.5}$ at 300 and 4 K. Small impurity peaks corresponding to Ba$_2$Fe$_4$S$_5$ are observed.}
	\label{fig2}
\end{figure}

\begin{figure}[t]
	\centering
	\includegraphics[scale=0.63]{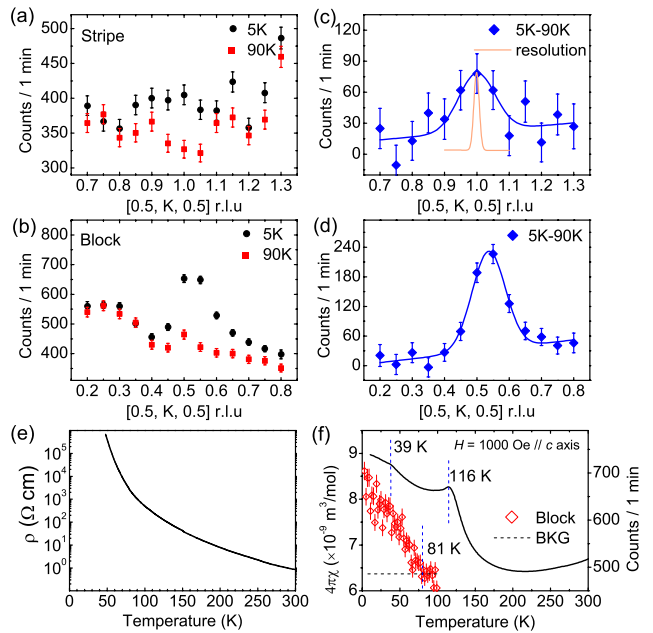}
	\caption{  Elastic scans along the $(0.5, K, 0.5)$ direction across (a) $K=1$ and (b) $K=0.5$ at 5 and 90 K. (c, d) The temperature differences between 5 and 90 K for $K=1$ and $K=0.5$, respectively. The solid lines are Gaussian fits to the data. The bars in (a-d) represent the statistical errors that correspond to one standard deviation. The instrumental resolution at $Q=(0.5, 1.0, 0.5)$ was determined by measuring the nuclear Bragg peak of $Q=(1, 2, 1)$. (e) Resistivity measured as a function of temperature. (f) Magnetic susceptibility measured with a 1000 Oe magnetic field parallel to the $b$ axis. The dashed lines mark kinks at 39 and 116 K. The open red diamonds show the temperature dependence of the block AF magnetic order parameter measured at $Q=(0.5, 0.5, 0.5)$. The order parameter reaches the background (BKG) at 81 K.
	}
	\label{fig2b}
\end{figure}

Figure \ref{fig2} shows NPD patterns for BaFe$_{2-\delta}$S$_{1.5}$Se$_{1.5}$ measured at 4 and 300 K. Small impurity peaks corresponding to Ba$_2$Fe$_4$S$_5$ were observed. No sharp magnetic peaks are seen in the NPD pattern at 4~K, in clear contrast to our previous measurements on as-grown powder samples with the same nominal composition, where there are strong magnetic peaks corresponding to the block AF order at low temperature\cite{yu2020structural}. The weak, broad feature centered around 2$\theta$ = 16.5$^{\circ}$ in Fig.~\ref{fig2}~(b) corresponds to block-type short-range magnetic order at $Q=(H, K, L)=(0.5,0.5,0.5)$. Here, $(H, K, L)$ are Miller indices for the momentum transfer $|Q|=2\pi\sqrt{(H/a)^2+(K/b)^2+(L/c)^2}$, where the lattice constants and refined structural parameters are listed in Table \ref {tab:table1}. The space group $Pnma$ agrees with previous reports. The structural refinement reveals the presence of $\sim$6\% iron vacancies, corresponding to a refined composition of BaFe$_{1.89(2)}$S$_{1.5(1)}$Se$_{1.5(1)}$. We note that the ratio of the S and Se atoms varies depending on the atomic site. The composition determined from EDS, normalized by the Ba content, is Ba$_{1.00(5)}$Fe$_{1.92(9)}$S$_{1.38(12)}$Se$_{1.37(15)}$.

\begin{figure}[t]
	\centering
	\includegraphics[scale=1.3]{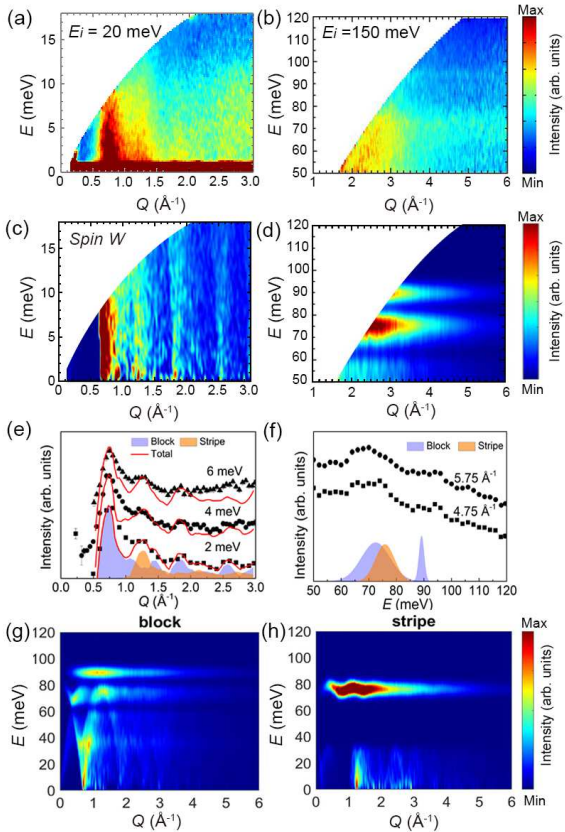}
	\caption{INS spectra $S(Q,\omega)$ for BaFe$_{1.89}$S$_{1.5}$Se$_{1.5}$ at 5~K with (a) \textit{E$_i$} = 20 and (b) 150~meV. Calculated $S_{SWT}(Q,\omega)$ using linear spin wave theory for (c) $E_i=20$ and (d) 150~meV with intensities from the block and stripe AF orders in a ratio of $4:1$.  The calculations include a convolution with the energy resolution of 1~meV and 7~meV for $E_i=20$~meV and $E_i=150$~meV, respectively. (e) Constant energy cuts through $S(Q,\omega)$ integrated over energy intervals of $2 \pm 0.5$, $4 \pm 0.5$, and $6 \pm 0.5$~meV using $E_i=20$~meV. The black dots are the experimental results and the red curves represent the simulation. (f) Constant momentum cuts through $S(Q,\omega)$ integrated over $4.75 \pm 0.25$ and $5.75 \pm 0.25$ \AA$^{-1}$. In (e) and (f), the shaded regions represent the calculated magnetic excitation intensities from the block and stripe AF orders as the same ratio in (c) and (d). (g) and (h) show the powder-averaged calculated spin excitation spectra for the block and stripe magnetic orders, respectively.}
	\label{fig3}
\end{figure}

To gain more insight into the short range magnetic structure, we conducted neutron diffraction measurements on a single crystal sample. The data shown in Fig.~\ref{fig2b} (a-d) reveal the appearance of broad magnetic peaks centered at $Q=(0.5, 1.0, 0.5)$ and $Q=(0.5, 0.5, 0.5)$, corresponding to the stripe and block AF orders, respectively. Considering that these magnetic peaks were not observable in the powder samples, instead requiring a large single crystal and the high flux at BT7 for unambiguous observation, the magnetic order is expected to be weak and short-range. The domain size of the stripe AF order along the leg direction is determined to be $\sim$47 \AA\ by convoluting the instrumental resolution. The temperature-dependent resistivity shown in Fig.~\ref{fig2b}(e) exhibits insulating behavior, consistent with the other compositions in this system\cite{Du2019,yu2020structural}. The temperature-dependent magnetic susceptibility in Fig.~\ref{fig2b}(f) displays kinks at 39 and 116 K, indicating multiple magnetic transitions. However, the neutron diffraction intensity measured at $Q=(0.5,0.5,0.5)$ (corresponding to the block AF order) shows a broad transition beginning at 81 K. These results suggest that the transition temperature may depend on the resolution and sensitivity of the instrument, as in a short-range cluster spin glass system\cite{Lu2014}.

To investigate the magnetic correlations present at low temperature, we conducted INS measurements on the same powder sample. Figures \ref{fig3} (a) and (b) display the INS spectra with incident energies of $E_i=20$ and 150 meV, respectively. Intense gapless magnetic excitations can be observed at low $Q$ in Fig. \ref{fig3} (a). Representative constant energy cuts at $E=2$, 4, 6 meV and constant $Q$ cuts at $Q=4.75$ and 5.75 \AA$^{-1}$ are plotted in Fig. \ref{fig3} (e) and (f). Two magnetic excitation modes centered at $Q = 0.75$ and 1.26 \AA$^{-1}$ can be identified in Fig. \ref{fig3} (e). The excitation mode at $Q = 0.75$ \AA$^{-1}$ is ascribed to the block AF order at the wave vector of (0.5, 0.5, 0.5)\cite{mourigal2015block}. The mode at $Q = 1.26$ \AA$^{-1}$ is consistent with the stripe AF wave vector at (0.5, 1.0, 0.5), such as that observed in BaFe$_2$S$_3$\cite{wang2017strong}. Thus, the spin excitations measured by INS demonstrate that both the block and stripe AF correlations exist in BaFe$_{1.89}$S$_{1.5}$Se$_{1.5}$. The magnetic excitations at higher energies in Fig. \ref{fig3} (b) and (f) show broad features, consistent with short-range magnetic correlations.

\begin{figure}[t]
	\includegraphics[scale=0.35]{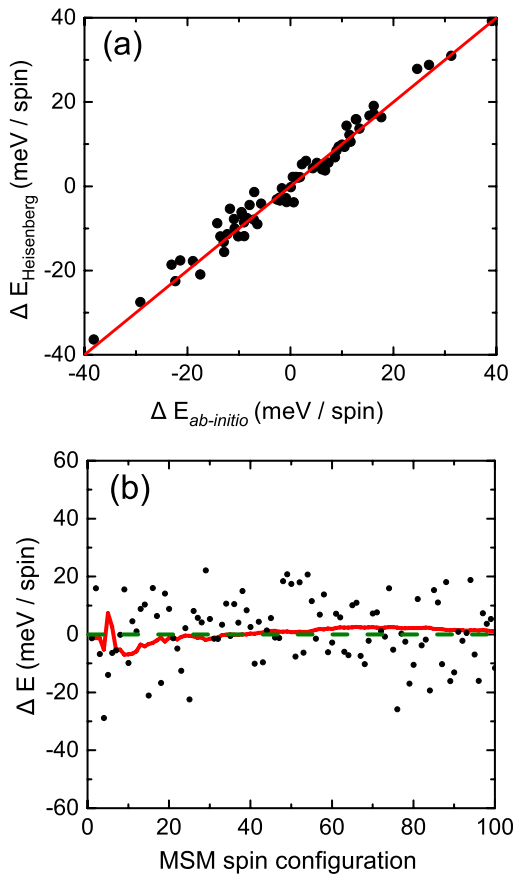}
	\caption{(a) The energies of 60 RGC spin configurations calculated via \textit{ab initio} methods versus the Heisenberg equation (Eq.~\ref{eq:1}) using the 12 fitted exchange interactions described in the main. The solid red line shows the equation $E_{Heisenberg} = E_{ab initio}$. (b) \textit{Ab initio}-calculated energy of 100 RGC spin configurations (black points) with the corresponding accumulated average (red solid line) compared to the average energy of the four spin configurations (green dashed line) used to approximate the PM phase.}
	\label{fig4}
\end{figure}

Using a classical Heisenberg model, we can describe the magnetic interactions in BaFe$_{1.89}$S$_{1.5}$Se$_{1.5}$ with the magnetic Hamiltonian
	\begin{equation}
	\begin{aligned}
	\textit{H$_{spin}$} = \sum_{i< j}J_{ij}\textbf{S}_i\ \cdot\ \textbf{S}_j,
	\end{aligned}\label{eq:1}
	\end{equation}	
where the sum runs over magnetic atoms in the compound and $J_{ij}$ are exchange integrals between Fe spins S$_i$ and S$_j$. To describe correctly the magnetic configurations of this system, both intraladder and interladder interactions should be considered. Here, we include 12 nonequivalent exchange interactions with bond lengths less than 7 {\AA}, labeled $J_1$ through $J_{12}$, which we illustrate in Fig. \ref{fig1} (a) and (b). Among the 12 $J$s, $J_1$-$J_7$ are intraladder interactions and the rest are interladder interactions.  Our choice of the 12 nonequivalent $J$s contains all the corresponding $J$s in Ref. \onlinecite{mourigal2015block}, which chose 8 exchange parameters to describe magnetic interactions in BaFe$_2$Se$_3$. The number of $J$s increases in the present system because of the lower symmetry. We determined the values of these interactions by fitting all 12 $J$ parameters to the magnetic energies calculated by DFT using 50  randomly generated collinear (RGC) magnetic configurations\cite{PhysRevB.91.165122,PhysRevB.67.134404}. The resulting values of $SJ_1$ through $SJ_{12}$ are -21.7, -24.6, 21.2, 25.5, 40.4, 9.3, 4.2, 0.5, 0.8, 3.8, 0.4, and 0.7 meV, where $S$ is assumed to be 2. Our calculated results are generally consistent with those in Ref. \onlinecite{mourigal2015block}. Specifically, to form a block magnetic state, $J_1$ and $J_3$ should be different in sign, which is consistent with the corresponding S$J_{1}$ and S$J_{1}^{\prime}$ in Ref. \onlinecite{mourigal2015block}. We also find that $SJ{_2} = 11$ meV and $SJ_{2}^{\prime} = 6$meV in Ref. \onlinecite{mourigal2015block}, with a similar ratio between the corresponding $J_4$ and $J_5$ interactions found in our results. It is worth noting that BaFe$_{2-\delta}$S$_{1.5}$Se$_{1.5}$ is a compound with a different atomic structure and symmetry than BaFe$_2$Se$_3$. It would be difficult to  compare $J$s between these two systems cleanly. However, given that the ground state of vacancy-free BaFe$_{2-\delta}$S$_{1.5}$Se$_{1.5}$ is the block magnetic phase observed in BaFe$_2$Se$_3$, it is reasonable to suppose that the exchange interactions in these two compounds could share some common features. To further check the completeness of our choice of $J$s, we performed calculations on another 60 RGC spin configurations. The fitted values for the 12 exchange integrals were then substituted into Eq. (\ref{eq:1}), and the energy, $E_{Heisenberg}$, was calculated for each of the 60 RGC spin configurations. In Fig.~\ref{fig4}(a), we compare $E_{Heisenberg}$ to the energies obtained from \textit{ab initio} calculations, E$_{\textit{ab initio}}$, using the same 60 RGC spin configurations. Each black circle represents one of the RGC configurations, and they all lie very close to the line $E_{Heisenberg} = E_{ab initio}$ shown in red. This close agreement demonstrates the completeness of our choice of exchange interactions in BaFe$_{1.89}$S$_{1.5}$Se$_{1.5}$.

With the calculated magnetic exchange integrals, we can simulate the spin wave spectra in linear spin wave theory using the $SpinW$ software package\cite{Toth2015}. The six most prominent magnetic exchange interactions, S$J_{1\cdots6}$, are included in the simulations of the powder-averaged magnetic spin excitation spectra of the block and stripe AF orders. The other calculated magnetic couplings are below 5~meV and would have negligible observable impact on the spin wave spectra. In Figs.~\ref{fig3} (c-f), we reproduce the spin wave spectra of BaFe$_{1.89}$S$_{1.5}$Se$_{1.5}$ by combining the block AF and stripe AF excitations with spectral weights of 80\% and 20\%, respectively. As shown in Fig. \ref{fig3} (e) and (f), the simulated results are in good agreement with the experimental data. The greater width of the observed spin excitations compared to the calculations results from the short-range magnetic order in the sample, as opposed to conventional long-range order. We note that theoretically one set of magnetic couplings would result in a degenerate magnetic ground state. Small deviations of the magnetic couplings could result in different magnetic orders. However, their effect on the magnetic excitation spectra may be negligible given a specific magnetic order.

\begin{figure}[t]
	\centering
	\includegraphics[scale=0.35]{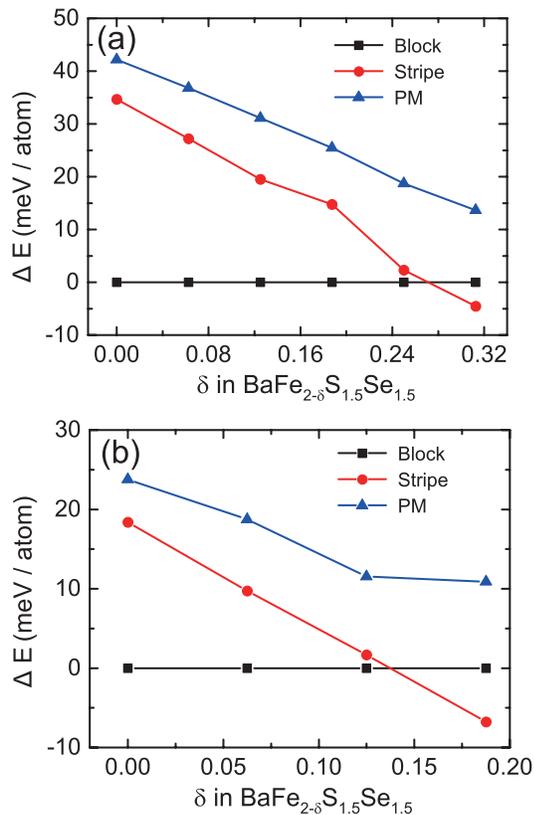}
	\caption{Variation of the block AF, stripe AF, and PM ground state energies (relative to the block AF energy) as a function of iron vacancy concentration without geometry optimization (a) and with geometry optimization (b).}
	\label{fig5}
\end{figure}

To explore the effect of iron vacancies on the magnetic properties, we performed DFT calculations for BaFe$_{2-\delta}$S$_{1.5}$Se$_{1.5}$ in the paramagnetic (PM) phase, block AF phase, and stripe AF phase with varying amounts of iron vacancies. First, we must define an appropriate model for the PM phase. Here we employed the magnetic sampling method (MSM). Within the Heisenberg model, a PM phase can be modeled by averaging multiple spin configurations in such a way that individual exchange interactions cancel, i.e., $\varphi = \sum_{k} \langle S_{i}(k)\cdot\ S_{j}(k)\rangle$ = 0, where $k$ denotes the spin configuration. This MSM method has been reliably used to model the PM phase of ZnCr$_2$O$_4$\cite{fennie2006magnetically} and CaBaCo$_4$O$_7$\cite{johnson2014cabaco}. By considering the 12 inequivalent exchange interactions, the PM phase in BaFe$_{2-\delta}$S$_{1.5}$Se$_{1.5}$ can be modeled with four collinear spin configurations resulting in vanishing exchange integrals between the Fe spins. To validate this approach, we performed calculations on 100 RGC spin configurations, whose average energy is representative of the energy of the PM phase. In Fig. \ref{fig4} (b), we plot the individual energies of the 100 RGC spin configurations (black dots) and their cumulative average (red curve) relative to the average energy of the four spin configurations selected for the PM model (green dashed line). The cumulative average energy of 100 RGC spin configurations converges to the energy from the four spin configurations representing the PM phase.

Having verified our model for the PM phase, we then calculated the total magnetic energy of BaFe$_{2-\delta}$S$_{1.5}$Se$_{1.5}$ in the block AF, stripe AF, and PM phases as a function of iron vacancy concentration.
Our theoretical calculations involve comparisons of energies alone. There is no entropy; that is, the calculations correspond to $T=0$ even for the PM phase. Figure \ref{fig5} (a) shows the results calculated using the experimentally refined parameters in Table \ref{tab:table1} without geometry and atomic position optimization. The total energy of the PM phase is always well above the other two phases and thus can be ruled out as the ground state. The block AF phase is more stable than the stripe AF phase when the Fe vacancy concentration is less than 12\%, above which the trend reverses. The introduction of Fe vacancies may affect the atomic configurations, so we also performed equivalent calculations with geometry and atomic position optimization, with the corresponding results shown in Fig. \ref{fig5} (b). In this case, the crossing point between the block AF and stripe AF is shifted to $\sim$6\%, which is very close to the vacancy concentration in the present sample.
Although it is not possible for us to fine tune the concentration of Fe vacancies in the DFT calculations due to the limitations of the supercell size and computational power, the calculated results clearly show that the competition between these two phases can be tuned by iron vacancies. Within a certain range of the vacancy concentrations around 6\%, the competition of these two phases may become so strong that the long range order is destroyed, consistent with the sample used in the present study. Microscopically, the disappearance of long-range magnetic order might also be caused by the dilution of exchange interactions due to the introduction of Fe vacancies, similar to the percolation effect in diluted magnetic semiconductors. However, such an effect typically requires a vacancy concentration well above 10\%. This dilution effect may explain the absence of the block AF order in severely iron deficient BaFe$_{1.8}$Se$_3$~\cite{Saparov2011}.

For a realistic sample, one would not expect the composition exactly locates at the magic concentration where the energies of the two magnetic orders are equal. As the temperature is lowered,  the regions with different Fe-vacancy contractions and therefore different preferred phases (block or stripe) would freeze out so that they are simply not able to come to equilibrium with each other.  Further, as the clusters freeze they will tend to exert random magnetic fields on each other thus enhancing the spin disorder and causing the time to equilibrate to diverge exponentially with decreasing temperature, forming a novel spin glass state with two short range orders. In other words, the disorder is ultimately a kinetic phenomenon with the system unable to find its true ground state whether it is stripe or block.

\section{Summary}
In summary, we have measured the short range magnetic order and spin excitation spectrum of coexisting short-range stripe and block AF order in the quasi-one-dimensional BaFe$_{1.89}$S$_{1.5}$Se$_{1.5}$. The iron vacancies are implicated in the lack of long-range magnetic order in the ground state and the existence of both stripe- and block-type excitation modes. First-principles calculations have confirmed the competition between stripe and block AF order, with maximum concentration expected around an iron vacancy concentration of $\sim$6\%, very close to the actual composition of the sample used in the present work. Using the calculated exchange interaction parameters, the simulated spin wave spectra match the measured results well. Our findings highlight the unusual sensitivity of the magnetic ground state to the presence of iron vacancies in the quasi-1D spin ladder system BaFe$_2$(S,Se)$_3$. These results are expected to be relevant to other pressure-induced superconductors in the family of quasi-1D iron pnictide and selenide systems.

\section{acknowledgments}

Work at Sun Yat-Sen University was supported by the National Natural Science Foundation of China (Grant No. 11904414, 12174454, 11904416, 11974432), the Guangdong Basic and Applied Basic Research Foundation (No. 2021B1515120015, 2019A1515011337), and National Key Research and Development Program of China (No. 2019YFA0705702, 2018YFA0306001, 2017YFA0206203). Work at University of California, Berkeley and Lawrence Berkeley National Laboratory was funded by the U.S. Department of
Energy, Office of Science, Office of Basic Energy Sciences, Materials Sciences and Engineering
Division under Contract No. DE-AC02-05-CH11231 within the Quantum Materials Program
(KC2202) and the Office of Basic Energy Sciences. A. D. Christianson was partially supported by the by the U.S. Department of Energy, Office of Science, Basic Energy Sciences , Materials Science and Engineering Division.
The experiment at Oak Ridge National Laboratory's Spallation Neutron Source was sponsored by the Scientific User Facilities Division, Office of Basic Energy Sciences, U.S. Department of Energy.
The identification of any commercial product or trade name does not imply endorsement or recommendation by the National Institute of Standards and Technology.

\bibliography{reference}
\end{document}